\documentclass[a4paper, 11pt]{article}
\usepackage{graphicx}
\usepackage{subcaption}
\usepackage{color}
\usepackage{enumitem}
\usepackage{amssymb}
\usepackage{amsmath}
\usepackage{times}
\usepackage{fullpage}
\usepackage{url}
\usepackage{wrapfig}
\usepackage{appendix}
\usepackage{xspace}
\usepackage{algcompatible}
\usepackage{algorithm}
\usepackage{algpseudocode}
\usepackage{listings}
\usepackage{verbatim}
\usepackage{multicol}
\usepackage{amssymb}
\usepackage{amsmath}
\usepackage{float}
\usepackage{caption}
\usepackage[numbers]{natbib}
\usepackage[a4paper, total={8.5in, 11in}, margin=1 in]{geometry}
\pdfoutput=1
\newcommand{\htab} {\emph{hash-table\space}}
\newcommand{\set} {\emph{set\space}}
\newcommand{\ins} {\emph{insert\space}}
\newcommand{\del} {\emph{delete\space}}
\newcommand{\find} {\emph{find\space}}
\newcommand{\lst} {\emph{list\space}}

\title{
\vspace{-1.5cm}
\bf Persistent Memory Programming Abstractions in Context of Concurrent Applications}
	\vspace{-1.5cm}\author{
		Ajay Singh \footnote{This work was done at Regal Lab of INRIA\&LIP6 during research internship.} \\
		\small cs15mtech01001@iith.ac.in \\
		\small IIT Hyderabad \\
		\and 
		Marc Shapiro \\
		\small marc.shapiro@acm.org\\
		\small INRIA \& LIP6 \\
		\and
        Gael Thomas\\
        \small gael.thomas@telecom-sudparis.eu\\
        \small Telecom SudParis\\
        }
\date{}
\begin{document}
\maketitle              
\thispagestyle{empty}
\abstractname{:
The advent of non-volatile memory (NVM) technologies like PCM, STT, memristors and Fe-RAM is believed to enhance the system performance by getting rid of the traditional memory hierarchy by reducing the gap between memory and storage. This memory technology is considered to have the performance like that of DRAM and persistence like that of disks. Thus, it would also provide significant performance benefits for big data applications by allowing in-memory processing of large data with the lowest latency to persistence. Leveraging the performance benefits of this memory-centric computing technology through traditional memory programming is not trivial and the challenges aggravate for parallel/concurrent applications. To this end, several programming abstractions have been proposed like NVthreads, Mnemosyne and intel's NVML. However, deciding upon a programming abstraction which is easier to program and at the same time ensures the consistency and balances various software and architectural trade-offs is openly debatable and active area of research for NVM community. 

We study the NVthreads, Mnemosyne and NVML libraries by building a concurrent and persistent set and open addressed hash-table data structure application. In this process, we explore and report various tradeoffs and hidden costs involved in building concurrent applications for persistence in terms of achieving efficiency, consistency and ease of programming with these NVM programming abstractions. Eventually, we evaluate the performance of the set and hash-table data structure applications. We observe that NVML is easiest to program with but is least efficient and Mnemosyne is most performance friendly but involves significant programming efforts to build concurrent and persistent applications.
}

\section{Introduction}
The advent of non-volatile memory (NVM) technologies is believed to enhance the system performance by getting rid of the traditional memory hierarchy by reducing the gap between \emph{memory} and the \emph{storage}. The recently proposed storage class memory (PCM, STT, Fe-RAM) sitting between memory and disk is supposed to have the performance like those of DRAM and persistence like those of disks\citep{lee2009architecting, huai2008spin}. NVM is byte addressable, allowing CPU to directly access the persistent storage much efficiently through load/store memory instructions. NVM would allow leveraging performance of RAM like memory which is persistent too. To quote Jim Pappas, SNIA, Vice Chairman- "convergence of storage and memory would be the most revolutionary change since the invention of transistors".

This opens up the intelligent question of  what should be the programming model to leverage the persistent memory? Such that it is easier to program applications and at the same time exploits the performance potential of NVM. From the perspective of system and application design, NVM programmers would require addressing various issues namely, \emph{consistency} due to \emph{cache re-ordering}, \emph{durability} due to volatile cache, building easily adaptable \emph{programming model} and \emph{garbage collection} of persistent memory objects\cite{pelley2014memory, venkataraman2011consistent,hsupractical}. These challenges are exacerbated in presence of multiple threads where the programming model needs to decide upon the global consistent state of the application which can be made durable and at the same time ensures per thread consistency of logs or metadata, if any.

To this end, many approaches have been proposed namely, Mnemosyne \cite{volos2011mnemosyne}, nvheaps \cite{coburn2011nv}, Atlas \cite{chakrabarti2014atlas}, Intel's NVML and NVthreads\cite{hsupractical}. However, the question of real gain in performance keeping in mind the trade-offs: overhead, scalability and correctness and ease of programming remain unanswered (especially if applications are multi-threaded) unless user himself writes applications. This motivated us to build some of the classical concurrent data structures in literature and test the programming abstraction libraries on the lines of above mentioned trade-offs as parameters.  

We focused on building concurrent and persistent \emph{set} \& \emph{hash-table} using the NVthreads, mnemosyene and NVML open source libraries. The \emph{set} application and closed addressed \emph{hash-table} both use coarse grained and fine grained list\cite{herlihy2011art} for the underlying implementation. Finally, we evaluated the performance of the applications and explored the trade-offs involved between ease of programming and efficiency. Our Major contribution is to explore the efforts required on part of the programmer to program a concurrent data-structure for persistence and mechanisms adopted by the libraries, which in turn will help the community to better understand the hidden issues of designing persistent/in-memory concurrent data structures and algorithms with NVM.
\section{Methodology \& Evaluation}
We implement a \set application and a closed addressed \htab which exports \emph{insert}, \del and \find operations. The two applications are \emph{concurrent and persistent}. Thus, the application upon restart appears to resume execution from the state it was in just before the crash. For comprehensive evaluation we build \emph{coarse-grained} and \emph{fine-grained} versions of the two applications.
Each list has nodes represented as $\left\langle key, value\right\rangle$ pair in increasing order of the keys. In coarse-grained version a lock has to be taken for the whole list before executing any operation upon it. In fine-grained list each operation acquires a pair of lock at corresponding region of the list which is similar to hand-over-hand locking mechanism.
Full source code can be found at: \textcolor{blue}{\url{https://github.com/aajayssingh/concurrent-persistent-datastructure}}

\textbf{Experiments with Mnemosyne:}
Mnemosyne\cite{volos2011mnemosyne} modifies kernel and have compiler instrumented code to support programming abstraction for the persistent data structure over NVM. It extends linux memory management module to virtualize NVM to enable multiple applications to execute over their own persistent memory segment. It uses \emph{scon} as a build tool instead of \emph{make} which might involve a small learning curve for the naive users. We used Mnemosyne's GCC version which requires lots of dependencies installation to start using the Mnemosyne for building the applications\cite{git-mnemo-gcc}.

Mnemosyne provides following primitives to program a persistent application:1) \emph{pmalloc:} to allocate the desired size of memory from the persistent heap of the NVM (here it is a mmaped file). 2) \emph{persistent:} to declare a typed variable with persistent keyword which implies that variable needs to be allocated from NVM.
 3) \emph{\_\_transaction\_atomic:} supports in-place update of data of more than 8 byte size. 4) \emph{pfree:} to release the persistent region allocated dynamically.

While writing the \set \& \htab applications underlying \emph{list} is declared as a global variable marked with MNEMOSYNE\_PERSISTENT keyword and all nodes are allocated persistent region from NVM using \emph{pmalloc}(). Synchronization is done using regular pthread mutex locks. To ensure failure atomicity\cite{hsupractical}, the modifications (insert \& delete) to the persistent nodes are made within lightweight durable transactions provided by Mnemosyne.

During recovery (after a crash or reboot) no explicit recovery code is required. The \lst which was declared statically (using MNEMOSYNE\_PERSISTENT keyword which is instrumented into compiler with section attribute) as a root object can be recursively traversed via its next pointer field to access whole list across multiple runs of the application.
One care though needs to be taken to reinitialize the mutex lock upon subsequent runs of the application. Otherwise, if a program crashes holding a lock, then same lock on the same node cannot be taken or program just may have indeterminate behaviour upon restart.

\textbf{Experiments with NVthreads:}
Nvthreads\cite{hsupractical} provides persistent programming abstraction for multi-threaded programs based on \emph{Dthreads}\cite{liu2011dthreads}. It exports following functions to program a persistent application: 1) \emph{nvmalloc:} takes size and an unique name of the variable to be allocated over NVM. 2)\emph{nvrecover:} to recover the memory allocated using the unique name provided during corresponding nvmalloc. 2)\emph{nvcheckpoint:} to flush all the cache resident data to the NVM.

It does not support persistent pointers, i.e. one cannot implicitly have a data structure containing a pointer field such that it is durable across restarts because the address to which the pointer points being a virtual address goes invalid once application exits. Though, there is a way to achieve this by explicitly maintaining the unique names of all the objects nvmalloced, and then manually reconstruct all the pointers during recovery using the unique names. For example, If we want to create a persistent linked list, we had to name each node uniquely and store this meta data either within the node or within a separate table in order to reconstruct the linked list during recovery. We tried to modify the recovery mechanism of the library to implement a persistent linked list without naming each node uniquely, but the approach has a drawback that during recovery, the order of the recovered nodes depends on the order the nodes were allocated. Thus, reconstructing the linked list in the state it was before crash was tricky and inefficient (because library records the location of each nvmalloced node in a mmaped file and nodes can be recovered only in order they were written on the file). So, keeping in mind our aim of exploring the library for implementing persistent concurrent data structures we decided to implement a linked list where each node will be nvmalloced with a unique string name. Our experiments with concurrent implementation show that NVML could not ensure consistent durability. Thus, we do not show it's evaluation in this manuscript however interested readers can find it in the report.

The underlying \lst of the \set and \htab is initialized using unique name (serves as root object to the persistent region) and the nodes are allocated with \emph{nvmalloc}. Each node contains its name, next node's name and a volatile next pointer. Thus, upon recovery each node of the \lst is recovered using \emph{nvrecover} and then volatile next pointers are again populated with the valid addresses of the recovered nodes. This results in significant size of the recovery code. The volatile next pointer is needed to avoid costly string read/write operations every time a node is dereferenced. Each thread which is a separate process infers durability semantics by intercepting lock-unlock pairs, and uses copy on write mechanism to buffer or commit modified memory pages to durable redo logs of NVthreads. It avoids ordering writes to NVM by just ensuring that writes to logs are ordered.

\textbf{Experiments with NVML}
Intel's NVML provides library support for programming persistent data structures without compiler/hardware modifications as a suite of software libraries. It simply makes excessive use of power of macros to implement type safety and transactional system for in-place updates to the underlying NVM device. The tutorials and online support available for the library along with its designs of API makes it more convenient to use and program.

We used C++ binding for \emph{libpmemobj} library. It provides following primitives to write the persistent application:1) \emph{make\_persistent:} to declare a typed variable meant to be allocated in NVM. 2) \emph{persistent\_ptr:} to declare a persistent pointer. 3) \emph{transaction::exec\_tx:} uses c++ lambda expression to perform in-place updates of data structures into the NVM. 4) \emph{delete\_persistent:} to delete an allocation from persistent region.

While writing the persistent application we needed to create a persistent object pool which is simply an abstraction for providing a memory mapped region using pool::create(). From this pool the application can allocate and deallocate its persistent regions throughout its lifetime using the root object returned by pool::create(). The underlying \lst can be operated using the primitives provided by the library. Each variable which is meant to be persistent is declared with keyword \emph{p} alongwith its type as in c++ style. And every pointer to persistent memory can be declared through smart pointer primitive i.e using \emph{persistent\_ptr}. The locks which are the member variable of \lst nodes are automatically re-initialized across multiple runs, unlike Mnemosyne. Thus, we do not need to worry about the lock states during recovery.

Durable transactions atomically modify the persistent data by maintaining a per thread undo log. Underlying architecture instructions CLFLUSH, CLWB ensure that data reaches persistent storage upon commit of the transactions. Each transaction uses significant amount of memory for logs and meta-data such that even 8MB of memory is not enough for a transaction with 3 nodes in the \emph{list}.
The \lst which was declared as a root object can be recursively traversed via its next pointer field to access whole list across multiple runs of the application. Thus no extra recovery code is needed. The program can simply start running with the \lst state as it was when crash occurred. The list state would be the one of last successful transaction which changed the state of the \emph{list}.

Table \ref{ptable} summarizes the different parameters we experimented with by building the concurrent and persistent \set and \emph{hash-table} by using Mnemosyne, NVML and NVthreads.
\begin{table}[h]
\centering
\scalebox{0.8}{
\begin{tabular}{l|lll}
Parameters                 & Mnemosyene    & NVML        & NVThreads   \\ \hline
Ease of programming        & Moderate      & Easy        & Moderate    \\
Crash consistency         &redo log            &undo log             &redo log             \\
Type safety                & NO            & YES         & NO      \\
Persistent pointers        & YES           & YES         & NO          \\
Kernel \& compiler support & YES           & NO          & NO          \\
Update tracking granularity         & Byte level & Page level  & Page level             \\
Recovery code overhead     & Minimal       & NO          & significant \\
Atomic updates             & Transactions  & Transaction & Locks      
\end{tabular}}
\caption{Observed parameters explored with \set \& \htab applications using the libraries.}
\label{ptable}
\end{table}
\vspace{-0.7cm}
\subsection{evaluation}\label{init-recover-run} 
\begin{figure}
 \begin{minipage}[h]{0.6\textwidth}
   \begin{subfigure}{\linewidth}
   \includegraphics[width=.5\linewidth]{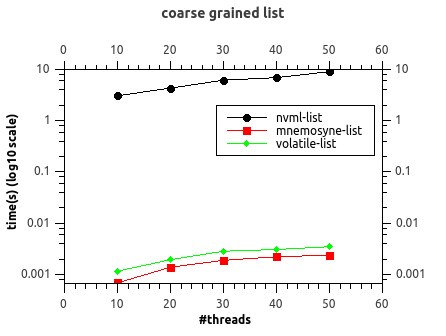}\hfill
   \includegraphics[width=.45\linewidth]{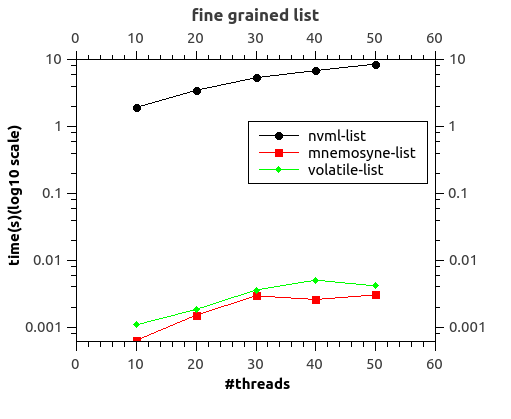}
   \caption{Time for coarse-grained \& fine-grained \lst.}
   \label{fig:a}
   \end{subfigure}
   \begin{subfigure}{\linewidth}
   \includegraphics[width=.49\linewidth]{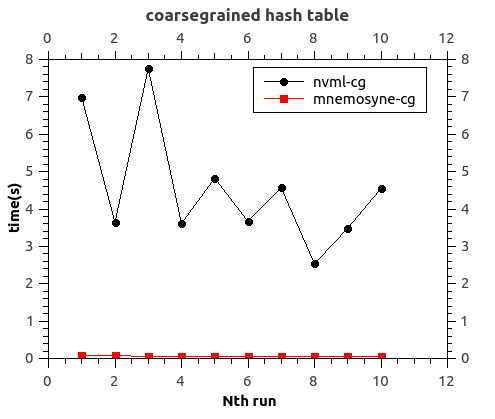}\hfill
   \includegraphics[width=.49\linewidth]{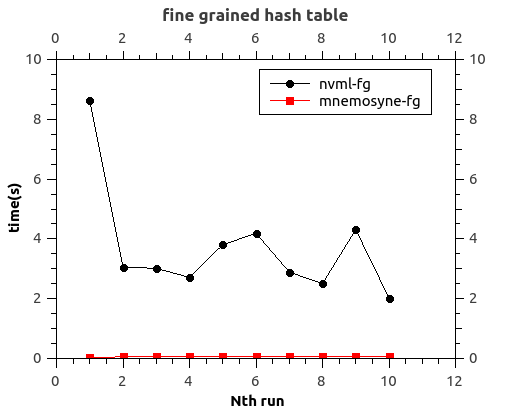}  
   \vfill
   \caption{Time for coarse-grained \& fine-grained \htab over\\10 run.}
   \label{fig:b}
   \end{subfigure}
 \end{minipage}\hfill
 \begin{minipage}[H]{0.4\textwidth}
   \begin{subfigure}{\linewidth}
   \vspace{-1em}\centering
   \includegraphics[scale=.36]{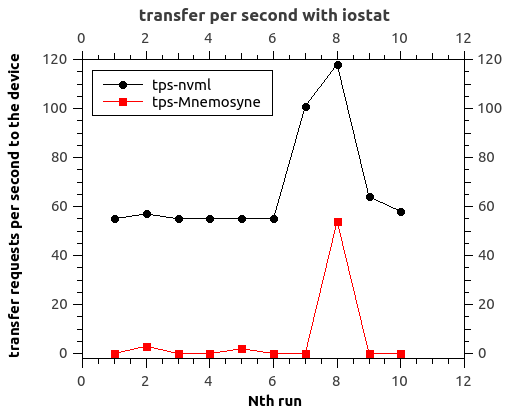}\caption{data transfers per second.}\label{fig:c}\vfill
   \includegraphics[scale=.36]{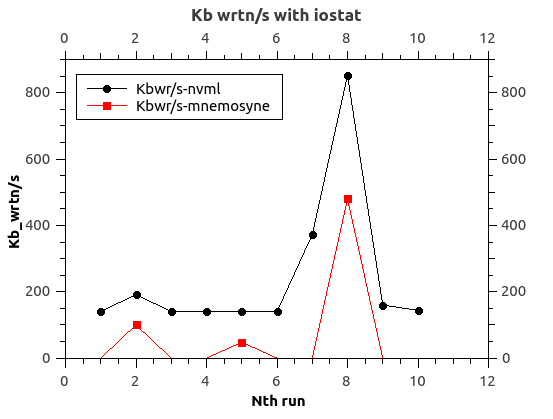}
   \caption{KBs of data written/sec.}
   \label{fig:d}
   \end{subfigure}
 \end{minipage}
 \caption{(a)\&(b) \set and \htab evaluation. (c)\&(d) analysis of time taken by NVML using iostat.}
 \end{figure}

\vspace{-0.2cm}
\textbf{Setup:} Performance of Mnemosyne, NVML and simple volatile \set is evaluated for coarse-grained as well as fine-grained version. We plot average wall clock time taken vs total number of threads (in Figure ~\ref{fig:a}). Each thread is designated to perform a single operation with 50\% of them are \ins{}operation, 40\% are \del{}operation and 10\% threads are \find{}operation. The applications were re-inited after every 2 runs. Thus, in first run the application intializes the underlying \lst of the \set and carries out all the operation and in second run the application carries out newer set of operation after recovering the previous \lst of the first run. While evaluating \htab(Figure ~\ref{fig:b}) we run the application multiple times each time recovering the previous \htab and applying new set of operation over this \emph{hash-table}. Thus, we plot average wall clock time vs $N_{th}$ run. Number of threads are 50. Rest of the evaluation setup remains same. All experiments were performed on x86\_64, Intel(R) Core(TM) i3-6100U CPU @ 2.30GHz system with 4 CPUs having Ubuntu GLIBC 2.23-0ubuntu9.\\ 
\\
\textbf{Analysis:} We observed that the \set implemented with Mnemosyne outperforms NVML and volatile implementations for both coarse-grained and fine-grained variants (Figure ~\ref{fig:a}). Similar trend follows for both the variants of closed addressed \htab(Figure ~\ref{fig:a}). 

\textbf{\small NVML performance analysis:}
Mnemosyne performs way faster than the NVML counterparts. This can be attributed to efficient logging and paging mechanism of the Mnemosyne implemented via compiler instrumentation. But, NVML has just focused on ease of programming aspect without caring much about the efficiency and optimization, to keep the library simple. Moreover, when we dig deeper, it is observed that NVML spends lot of time in doing IO operation in comparison to Mnemosyne. We validate this through \emph{iostat} utility of linux to plot number of KB written per second and number of transfers requested per second to the underlying  device for the applications running with NVML and Mnemosyne (Figure ~\ref{fig:c} \& Figure ~\ref{fig:d}). \emph{Iostat} utility is used for coarse-grained \lst only as this gives enough insight for the behaviour of other application variants as well.   
\section{Conclusion \& Future Work}
Mnemosyne applications worked correctly and its performance (as can be seen from the plots) is the best amongst all the libraries subjected to the study. However, it can be attributed much to the compiler specific changes and kernel modifications. In the case of Mnemosyne, ensuring type safety of the persistent data structures is complex and may be error prone as we observed that freeing the deleted node's memory failed due to typecasting run time error. Also, it did not support garbage collection. We had to write the minimal recovery code specifically to re-initialize the locks taken, if any, before the crash of the application. It involves setting up the system by installing dependencies, uses \emph{scon} as the build tool and no type safety mechanism is provided which causes us to categorize it as moderate with respect to ease in programming. 

NVthread's approach was most tricky to program with as it did not support the persistent pointers. The alternative implementation of fine-grained and coarse-grained lists programmed through it failed to recover some of the nodes. This, we observed occurred because the recovery meta-data got corrupted. On deeper analysis we found that NVthreads infers the durability semantics using the synchronization points (lock unlock points)\cite{chakrabarti2014atlas} of the application. It is not clear how a fine-grained list like the one with hand over hand locking (which involves a complex nested pairwise locking) could be built efficiently, as the NVthreads approach treats nested locks equivalent to one single critical section. Moreover, It is known that lock based programs are buggy and difficult to program. However, the open addressed hash table worked correctly as it only involved a contiguous memory allocation without the use of any pointer member variable and used simple coarse-grained locks for synchronization. NVthreads also required the significant amount of recovery code and the garbage collection did not work for the deleted nodes. Moreover, it does not prevent volatile pointer to point to a non-volatile memory address which can be detrimental to application correctness.

On the other hand NVML though being very slow in terms of performance was easiest to program with and required no recovery code for the two concurrent list variants. Mnemosyne has been the quickest of all as it implements own virtual paging mechanism and instruments compiler to implement the persistency primitives, also it implements an efficient logging mechanism for its durable transaction to update data atomically, which reduces performance cost significantly. NVML (as a design decision) has completely focused on ease of programming the application and implements all the features at the software level in form of libraries, with the heavy dependency on macros. NVML's durable transactions, in fact, uses MBs of metadata and uses hardware persistence primitives multiple times to flush data to NVM device leading to its slow performance.

Thus, with respect to ease in programming we can rate NVML, Mnemosyne and NVthreads as easy, moderate and difficult respectively and with respect to the performance we observe NVML significantly lags behind Mnemosyne.  We plan to utilize the insights gained with these experiments to implement a programming abstraction for NVM such that it balances the trade-offs and optimizes the system throughput, especially for concurrent applications. The complete source code of the experiments and detailed report of the work is available at \url{https://github.com/aajayssingh/concurrent-persistent-datastructure}.
\newline

\end{document}